\documentclass[aps,prd
,preprint,tightenlines,nofootinbib,showpacs
]{revtex4}
\usepackage{amssymb,latexsym}
\usepackage{amsmath,amsbsy,bbm}
\usepackage{epsfig,bm}
\usepackage{graphicx,comment}
\DeclareMathOperator{\tr}{tr}

\begin{document}
\def\a{{\alpha}}
\def\b{{\beta}}
\def\d{{\delta}}
\def\D{{\Delta}}
\def\e{{\varepsilon}}
\def\g{{\gamma}}
\def\G{{\Gamma}}
\def\k{{\kappa}}
\def\l{{\lambda}}
\def\L{{\Lambda}}
\def\m{{\mu}}
\def\n{{\nu}}
\def\o{{\omega}}
\def\O{{\Omega}}
\def\S{{\Sigma}}
\def\s{{\sigma}}
\def\th{{\theta}}

\def\ol#1{{\overline{#1}}}

\def\Dslash{D\hskip-0.65em /}
\def\Dtslash{\tilde{D} \hskip-0.65em /}
\def\sumint{\sum \hskip-1.35em \int_{ \hskip-0.25em \underset{\bm{k}}{\phantom{a}} } \, \, \,}

\def\CPT{{$\chi$PT}}
\def\QCPT{{Q$\chi$PT}}
\def\PQCPT{{PQ$\chi$PT}}
\def\tr{\text{tr}}
\def\str{\text{str}}
\def\diag{\text{diag}}
\def\order{{\mathcal O}}

\def\meff{{m^2_{\text{eff}}}}

\def\Meff{{M_{\text{eff}}}}
\def\cF{{\mathcal F}}
\def\cS{{\mathcal S}}
\def\cC{{\mathcal C}}
\def\cE{{\mathcal E}}
\def\cB{{\mathcal B}}
\def\cT{{\mathcal T}}
\def\cQ{{\mathcal Q}}
\def\cL{{\mathcal L}}
\def\cO{{\mathcal O}}
\def\cA{{\mathcal A}}
\def\cR{{\mathcal R}}
\def\cH{{\mathcal H}}
\def\cW{{\mathcal W}}
\def\cM{{\mathcal M}}
\def\cD{{\mathcal D}}
\def\cN{{\mathcal N}}
\def\cP{{\mathcal P}}
\def\cK{{\mathcal K}}
\def\Qt{{\tilde{Q}}}
\def\Dt{{\tilde{D}}}
\def\St{{\tilde{\Sigma}}}
\def\cBt{{\tilde{\mathcal{B}}}}
\def\cDt{{\tilde{\mathcal{D}}}}
\def\cTt{{\tilde{\mathcal{T}}}}
\def\cMt{{\tilde{\mathcal{M}}}}
\def\At{{\tilde{A}}}
\def\cNt{{\tilde{\mathcal{N}}}}
\def\cOt{{\tilde{\mathcal{O}}}}
\def\cPt{{\tilde{\mathcal{P}}}}
\def\cI{{\mathcal{I}}}
\def\cJ{{\mathcal{J}}}

\def\eqref#1{{(\ref{#1})}}
\preprint{UMD-40762-441}

\title{Time Dependence of Nucleon Correlation Functions in Chiral Perturbation Theory}

\author{Brian C.~Tiburzi}
\email[]{bctiburz@umd.edu}
\affiliation{Maryland Center for Fundamental Physics, Department of Physics, University of Maryland, College Park,  MD 20742-4111, USA}

\date{\today}


\begin{abstract}
We consider corrections to nucleon correlation functions arising from times that are far from the asymptotic limit. 
For such times, 
the single nucleon state is contaminated by the pion-nucleon and pion-delta continuum. 
We use heavy baryon chiral perturbation theory to derive the spectral representation of the nucleon two-point function. 
Finite time corrections to the axial current correlation function are also derived. 
Pion-nucleon excited state contributions drive the axial correlator upward, 
while contributions from the interference of pion-delta and pion-nucleon states drive the axial correlator downward. 
Our results can be compared qualitatively to optimized nucleon correlators calculated in lattice QCD,
because the chiral corrections characterize only low-energy excitations above the ground state. 
We show that improved nucleon operators can lead to an underestimation of the nucleon axial charge. 
\end{abstract}

\maketitle

\section{Introduction}

Lattice gauge theory simulations continue to make impressive progress towards
addressing quantitatively the non-perturabative regime of QCD~\cite{DeGrand:2006aa}. 
Two decades ago, it was thought impossible that lattice QCD simulations would
confront experimental data without orders of magnitude of increased computing power,
and considerable algorithmic advances. 
Today, however, state-of-the-art simulations have pushed forward on all fronts by:
including fully dynamical quarks, shrinking the lattice spacing, enlarging the lattice 
volume, and decreasing the size of the light quark masses 
(including a sprint to the physical point~\cite{Aoki:2008sm}). 
The current status of these simulations is reviewed in~\cite{Jansen:2008vs}. 
Lattice QCD is approaching a point of not only complementing existing experimental programs, 
but potentially guiding future ones. 
A thorough overview of hadron structure from recent and forthcoming lattice calculations is presented in~\cite{Zanotti:2008zm}.

Performing lattice QCD calculations with light pions is an exciting recent development, 
but one that is accompanied by new problems. 
At fixed lattice sizes, for example, finite size effects from virtual pion fluctuations will become increasingly important. 
Another notorious issue in dealing with hadrons
on the lattice is the signal-to-noise problem. 
In the chiral regime, this problem will become rather acute. 
Consider the nucleon two-point correlation function, $G(\tau)$.
Over long times, $\tau$, the two-point function has an exponential 
falloff governed by the nucleon mass, $M_N$.
Over such long times, however, the statistical noise in the correlator, $\Sigma(\tau)$, 
is dominated by its coupling to three-pions~\cite{Lepage:1989hd}.
This leads to the signal-to-noise ratio having the behavior
\begin{equation}
\frac{ G(\tau) }{ \Sigma(\tau)}
\overset{\tau \gg 1}{\sim}
\sqrt{N} 
\exp \left[ - \left( M_N - \frac{3}{2} m_\pi \right) \tau \right]  
\notag
,\end{equation}
with $N$ as the number of independent measurements on gauge configurations. 
As pion masses enter the chiral regime, the long-time 
behavior of the correlation function is dominated by 
statistical noise. 
Generally this limits lattice QCD measurements of hadronic 
correlation functions to times that are not ideally long.%
\footnote{
Excluding pion zero modes by using parity-orbifold boundary conditions
has been suggested as a means to overcome the signal-to-noise problem~\cite{Bedaque:2007pe,Bedaque:2008hn}.
}

In this work, 
we consider corrections to hadronic correlation functions arising from the pion continuum. 
The Lehmann-Symanzik-Zimmerman (LSZ) reduction formula provides the 
field theoretic recipe for producing hadronic states in the limit of asymptotically long times. 
Over shorter times, 
however, 
there are transient fluctuations to multiparticle states within a hadron.
As the signal-to-noise problem restricts lattice QCD simulations to times far from the asymptotic regime where LSZ applies,  
we are motivated to investigate corrections to the reduction formula. 
To focus our discussion, 
we consider the chiral dynamics of the nucleon.
We derive corrections to nucleon two- and three-point functions using chiral perturbation theory.
In heavy baryon chiral perturbation theory, 
one expands about small energies above the nucleon mass. 
The nucleon operator should thus be thought of as some baryon interpolating field
that has been highly optimized through the removal of high-lying modes, 
i.e.~modes $k$ that are
$k \gtrsim \L_\chi$ 
above the nucleon mass, 
where 
$\L_\chi \sim 1 \, \texttt{GeV}$ 
is the chiral symmetry breaking scale.  
In the language of lattice QCD, 
the renormalized nucleon operator in chiral perturabtion theory 
behaves similarly to a quark-smeared baryon interpolating field. 
It is obviously difficult to compare our results quantitatively with those from similarly optimized nucleon interpolating fields currently used in lattice QCD studies.
Our results, however, provide a clean field-theoretic example of the difficulties encountered from excited state contamination. 
Additionally qualitative information, 
such as the sign of corrections from excited states, 
may be garnered from our computations. 
For three-point functions, 
the sign is non-trivial due to the lack of a spectral representation. 
We find that contributions from pion-nucleon states are driving  the axial correlator up, 
while those from the interference of pion-delta and pion-nucleon states drive the correlator down.

The organization of our presentation is as follows. 
First in Section~\ref{s:Spectral},
we derive the spectral representation of the nucleon propagator in heavy baryon chiral perturbation theory. 
Here we investigate the effects from pion fluctuations on the time dependence of nucleon two-point functions.
Next in Section~\ref{s:ThreePoint}, 
we extend our consideration to the LSZ reduction of the nucleon three-point function involving the axial-vector current. 
Sizable corrections to the axial correlator are found, 
and can lead to overestimation of the nucleon axial charge, 
$G_A$.
On the other hand, 
we describe a scenario in which the axial charge can be underestimated by improving the nucleon overlap in two-point functions. 
Concluding remarks are given in Section~\ref{s:summy}.

\section{Spectral Representation}
\label{s:Spectral}

Consider a free fermion field, 
$N(x)$, 
of mass 
$M$. 
In the limit of large mass, 
the velocity becomes a good quantum number in coordinate space; 
and, 
to leading order in 
$1/M$, 
we can write down a heavy fermion effective Lagrangian~\cite{Georgi:1990um,Jenkins:1990jv,Jenkins:1991es}, 
$\cL =N^\dagger  D_4 N$,
where $N$ is a two-component Pauli spinor. 
The coordinate space propagator, 
$D(x,0)$,
for the free heavy fermion has a simple form,
$D(x,0) =
\theta( x_4) \delta ( \bm{x} )
e^{ - M x_4 }$.
When interaction terms are included in the Lagrangian, 
the full two-point function,
$G(x,0)$,
has the general form
\begin{eqnarray}
G(x,0) 
=
\theta( x_4) \delta (  \bm{x} )
e^{ - M  x_4 }
\int_0^\infty dE 
\, \rho(E) e^{ - E x_4 }
,\end{eqnarray}
which follows from taking the large mass limit of the 
K\"all\'en-Lehmann spectral representation. 
The spectral function, $\rho(E)$, is a positive
function in the distribution sense. 
Assuming there is an isolated single-particle state
corresponding to the 
$N$, 
we write
$\rho(E) = \delta(E) + \ol \rho(E)$.
The residual spectral function, $\ol \rho(E)$,
is then assumed to vanish below the energy 
$E_{\text{th}}$, 
with $E_{\text{th}} > 0$. 
With the single-particle state isolated, 
the spectral representation has the form
\begin{eqnarray}
G(x,0) 
&=&
\theta( x_4) \delta ( \bm{x} )
e^{ - M  x_4 }
\left[
1
+ 
\int_{E_{\text{th}}}^\infty
dE \, 
\ol \rho(E)
e^{ - E  x_4}
\right]
.\end{eqnarray}
In the absence of bound states, 
$E_{\text{th}}$ 
is the threshold energy to create a multiparticle state.

Suppose there is an excited state, $N^*$, of mass $M^*$ contributing to the spectral function.
The spectral weight around the value $E = M^* - M$ then contains a delta-function.
For asymptotically large time separation between source and sink, 
$x_4 \gg 1$, 
this excited-state contribution is suppressed relative to the ground state by an exponentially small factor:
$\exp [ - ( M^* - M)  x_4]$.
The contributions from multiparticle states too are suppressed in the limit of asymptotic time separation. 
If we assume the spectral weight near threshold is of the form
$\ol \rho(E) \propto  ( E - E_{\text{th}} )^{n-1}$,
then the contribution from the multiparticle branch-cut is suppressed by a factor of
$(x_4)^{-n}
\exp ( - E_{\text{th}} x_4 )$,
for large time separation. 
The power $n$ is determined by the available phase space at threshold, 
and the required angular momentum for multiparticle production. 
If correlation functions are not deduced at large enough time separations,
the light pions will likely present difficulty for bound states in QCD, because 
$E_{\text{th}} = m_\pi$.

\subsection{Chiral Computation}

Treated as heavy fermions, 
free nucleons can be described the heavy fermion Lagrangian,  
with 
$N$ 
now upgraded to a two-component isospinor, 
$N = ( p, n )^T$. 
Interactions between pions and nucleons; 
as well as pions, nucleons and deltas
can be included in a way consistent with chiral symmetry~\cite{Bernard:1992qa,Hemmert:1997ye}. 
These interactions, 
moreover, 
can be systematically organized in terms of a small expansion parameter 
$\varepsilon$, 
where the quantities:
$k / M$ with 
$k$ 
as residual baryon momentum,  
$p / \L_\chi$ 
with 
$p$ 
as pion momentum and 
$\L_\chi = 4 \pi f$ 
as the chiral symmetry breaking scale, 
$m_\pi / \L_\chi$,  
and 
$\D / \L_\chi$ 
with 
$\D$ 
as the nucleon-delta mass splitting, 
are all treated to be of the size 
$\varepsilon$.  
The leading interactions among the baryons and pions 
are at 
$\cO(\varepsilon)$, 
and contained in the Lagrangian
\begin{eqnarray} \label{eq:Lint}
\cL 
=
2 g_A  \, 
N^\dagger 
\bm{S} \cdot \bm{A}
\, N
+ 
g_{\D N}
\left[
\bm{T}^\dagger  \cdot \bm{A} \,  N
+ 
N^\dagger \bm{A} \cdot \bm{T}
\right]
.\end{eqnarray}
Here $\bm{S} = \bm{\sigma} / 2$ is the spin operator,
and the quartet of deltas are packaged in a Rarita-Schwinger field 
$\bm{T}$. 
The axial-vector field of pions is at leading order
$\bm{A} = \frac{1}{f} \bm{\nabla} \phi + \ldots$, 
with $\phi = \tau^a \pi^a$.

%
\begin{figure}
\epsfig{file=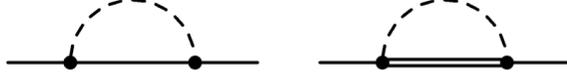,angle=270,width=7.5cm}
\caption{\label{f:NM} 
One-loop diagrams contributing to the nucleon two-point function. 
Single (double) lines denote nucleons (deltas), while the 
dashed lines denote pions. The filled circles denote 
axial couplings from the interaction Lagrangian, Eq.~\eqref{eq:Lint}.  
}%
        \end{figure}
%

Using Eq.~\eqref{eq:Lint}, 
we can determine the interacting nucleon two-point function
at one-loop order in the chiral expansion and thereby determine
the spectral function.
The contributing diagrams are shown in Figure~\ref{f:NM}. 
 Their computation is complicated by divergences which 
 we regulate using dimensional regularization. 
To renormalize the two-point function, we choose to work in 
coordinate space.%
\footnote{
It is instructive to carry out the computation additionally in momentum space. 
After the mass and wavefunction renormalization have been taken into account, 
the renormalized self-energy 
$\Sigma^{\text{ren}}(E)$ 
has the form
\begin{eqnarray} \notag
\Sigma^{\text{ren}}(E) = \Sigma(E) - \Sigma(0) - E \frac{d}{dE} \Sigma(0)
,\end{eqnarray}
but still contains divergences. These divergences only contribute to the
extreme short-time behavior of the two-point function, i.e.~contributions
to 
$G(\tau)$ 
in Eq.~\eqref{eq:G} of the form 
$\delta'(\tau)$, 
and 
$\delta''(\tau)$.
The spectral function, however, can be deduced from the regulated self-energy
from analyticity. 
The relation
\begin{eqnarray} \notag
\rho(E) = - \frac{1}{\pi} \Im \mathfrak{m} \left[ G(E) \right]
,\end{eqnarray}
produces the same spectral function as deduced in coordinate space.
 } %
This is particularly advantageous for our 
consideration of the LSZ reduction for three-point functions which we undertake in Section~\ref{s:ThreePoint}.

We first define the zero-momentum projected two-point function
\begin{eqnarray} \label{eq:G}
G(\tau) \equiv \int d\bm{x} \, G(\bm{x}, \tau; \bm{0}, 0 )
,\end{eqnarray}
where we have used spacetime translational invariance to locate 
the nucleon source at the origin of our coordinate system. 
To renormalize the zero-momentum two-point function, we require
that there be a single nucleon contribution at asymptotically large $\tau$. 
This consists of two parts, 
the mass renormalization
\begin{eqnarray}
\lim_{\tau \to \infty}
\left[
- \frac{d}{d\tau} \log G(\tau) 
\right]
\equiv 
M_{\text{phys}}
,\end{eqnarray}
that fixes the nucleon mass to its physical value 
(which thus absorbs its pion mass dependence), 
and the wavefunction renormalization
\begin{eqnarray}
\lim_{\tau \to \infty}
\left[
e^{M_{\text{phys}} \tau} G(\tau)
\right]
\equiv 
1
,\end{eqnarray}
that fixes the single nucleon probability to unity. 
Carrying out the one-loop chiral computation of the nucleon two-point function, we find%
\footnote{
It is straightforward to derive analogous results in a finite spatial volume. 
With periodic boundary conditions, 
the lattice momentum modes are quantized in the form
$\bm{k} = 2 \pi \bm{n} / L$, 
where 
$\bm{n} \in \mathbb{Z}^3$, 
and
$L$ is size of the lattice, 
which is assumed to be the same in each of the three spatial directions.
Integrals over the energy are replaced by sums over the momentum modes
\begin{equation} \notag
\int dE \, E \, f(E) 
\to 
\sum_{\bm{n}} 
\frac{2 \pi n}{L} 
f\big(  \sqrt{\bm{k}^2 + m_\pi^2} \big) 
,\end{equation}
where 
$n = \sqrt{\bm{n}^2}$. 
We are assuming a large enough volume, 
$m_\pi L \gg 1$,
so that the sum can be replaced by the integral for which the spectrum is continuous. 
Finite volume corrections to this approximation can be calculated as the difference 
between the sum and integral. 
}
\begin{eqnarray} \label{eq:HBspectral}
G(\tau) = \theta(\tau) e^{- M_{\text{phys}} \tau} 
\left[
1
+
\int_{m_\pi}^\infty dE \, 
\ol \rho_{\pi N}(E) 
e^{ - E \tau}
+ 
\int_{m_\pi + \D}^\infty dE \, 
\ol \rho_{\pi \D} (E) 
e^{ - E \tau}
\right]
,\end{eqnarray}
where the pion-nucleon and pion-delta 
fluctuations are described the by spectral weights
\begin{eqnarray}
\ol \rho_{\pi N}(E) 
&=&
\frac{6 g_A^2}{( 4 \pi f)^2}
\frac{[E^2 - m_\pi^2]^{3/2}}
{E^2}, \label{eq:Npi}
\\
\ol \rho_{\pi \D} (E) 
&=&
\frac{16 g_{\D N}^2}{3 ( 4 \pi f)^2}
\frac{[(E- \D)^2 - m_\pi^2]^{3/2}}
{E^2} \label{eq:Dpi}
.\end{eqnarray}


For asymptotically large times, 
the decaying exponential from the single nucleon state is the dominant contribution to the two-point function. 
Modification from pion interactions can be deduced in this limit by considering the spectral weights near threshold. 
The pion-nucleon and pion-delta spectral functions both vanish as the 
$3/2$-power 
of the energy available at threshold, 
$\propto ( E - E_{\text{th}})^{3/2}$.
This power-law behavior is due to the two-body phase space, 
$\propto ( E - E_{\text{th}})^{1/2}$, 
and the requirement that the pion and baryon be in a relative 
$p$-wave. 
Because the pion-delta threshold is larger than the pion-nucleon threshold by the mass splitting 
$\D$, 
the pion-delta contributions are suppressed relative to pion-nucleon contributions by the exponential
factor
$\exp ( - \D \tau )$.
Hence we shall neglect the delta contribution in asymptopia.

Expanding the pion-nucleon spectral weight about threshold, 
we arrive at the asymptotic expansion of the nucleon two-point function
\begin{eqnarray}
G(\tau) 
\overset{\tau \gg 1}{\longrightarrow}
e^{ - M_{\text{phys}} \tau}
\left[
1
+ 
\sqrt{2 \pi} \, 
\left(
\frac{3 g_A m_\pi}{4 \pi f}
\right)^2
\frac{ e^{ - m_\pi \tau} }{(m_\pi \tau)^{5/2}}
\left(
1 
- 
\frac{25}{8}
\frac{1}{m_\pi \tau}
+ 
\frac{1785}{128}
\frac{1}{( m_\pi \tau )^2}
+
\ldots
\right)
\right]
.\notag \\
\label{eq:asymp}
 \end{eqnarray}
The $\ldots$ denotes power-law suppressed terms proportional to $1 /  m_\pi \tau $. 
The expansion for large times is only an asymptotic one,
which is evidenced by the particularly large numerical coefficients of higher-order terms. 
The times over which the expansion in $1/m_\pi \tau$ results in a controlled
approximation to Eq.~\eqref{eq:HBspectral} are quite large, $\tau \gtrsim 50 \, a$, for $a = 0.125 \, \texttt{fm}$.
For smaller times, the first term in the expansion yields the best agreement with Eq.~\eqref{eq:HBspectral}, 
as is characteristic of asymptotic expansions used outside their range of validity~\cite{Tiburzi:2008ma}. 
To analyze nucleon two-point functions, however, we shall not use the approximation in Eq.~\eqref{eq:asymp}, 
but return to the full form in Eq.~\eqref{eq:HBspectral}.

\subsection{Mass Extraction}

%
\begin{figure}
\epsfig{file=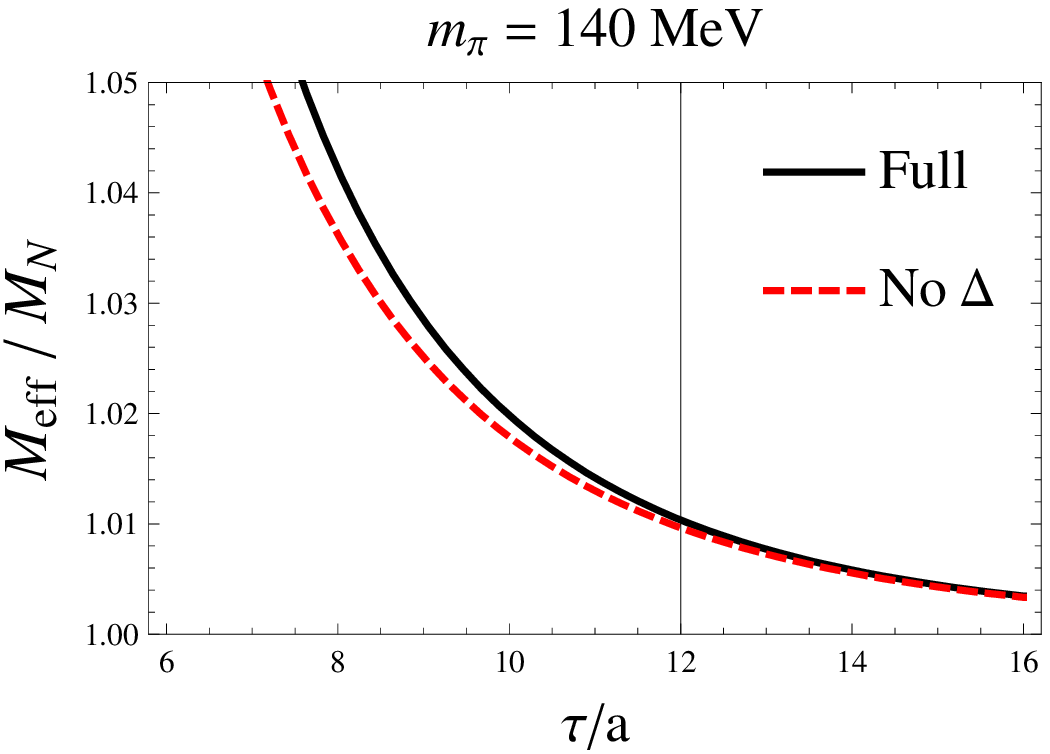,width=7.5cm}
$\quad$
\epsfig{file=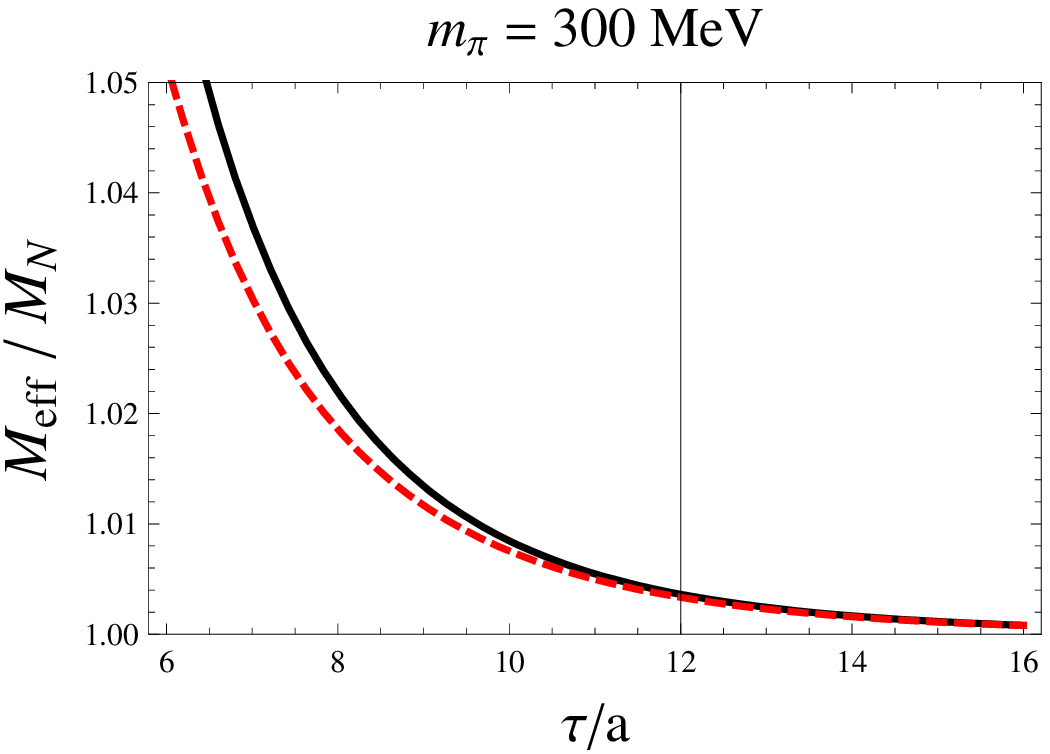,width=7.5cm}
\caption{\label{f:Mass} 
Interacting nucleon effective mass plots. 
The curve denoted by \emph{Full} corresponds to the effective mass derived from the fully interacting two-point function in Eq.~\eqref{eq:HBspectral}, 
while that denoted by \emph{No $\D$} is the same two-point function evaluated without delta resonance contributions, i.e.~$g_{\D N} = 0$. 
The vertical bar symbolizes the region where we imagine noise might dominate the correlation function.
}%
        \end{figure}
%

Now we consider the effect of the pion continuum on the nucleon two-point function. 
A typical quantity used for guiding the eye in fits of lattice two-point functions is the effective mass, 
and we shall employ this tool to investigate the interacting nucleon two-point function derived 
from chiral perturbation theory. 
The effective mass function is defined as
\begin{equation} \label{eq:Meff}
M_{\text{eff}}(\tau / a)
=
- \log 
\frac{ G( \tau + a) }{G(\tau)}
.\end{equation}
We employ lattice units using a temporal lattice spacing of
$a = 0.125 \, \texttt{fm}$, 
which is the spacing in some current-day lattices. 
For large $\tau / a$, the effective mass should become flat
with a value corresponding to the nucleon mass. 
The onset of a plateau in the effective mass gives one an indication of
when the ground state dominates the two-point correlation function. 
Of course, 
in actual lattice calculations the signal-to-noise problem for the nucleon limits one to times that are not ideally long.

In Figure~\ref{f:Mass}, 
we make two effective mass plots using
the interacting nucleon two-point function in Eq.~\eqref{eq:HBspectral}. 
The low-energy constants are fixed to their known values, 
specifically we use
$g_A = 1.27$, 
$g_{\D N} = 1.5$, 
$\D = 290 \, \texttt{MeV}$, 
and 
$f = 130 \, \texttt{MeV}$. 
We plot the ratio of the effective mass to the nucleon mass for two values
of the pion mass, 
$m_\pi = 140$, and $300 \, \texttt{MeV}$. 
The onset of a plateau in the effective mass hence corresponds the ratio 
$M_{\text{eff}} /M_N$ 
approaching unity
which is exhibited in both panels of the figure. 
We see, moreover, that the pion-delta continuum gives negligible
contributions due to the exponential suppression factor, $\exp ( - \Delta \tau )$. 
In a recent lattice study~\cite{WalkerLoud:2008bp}, 
the noise begins to dominate the signal around 
$\tau / a = 12$ 
at 
$m_\pi \sim 300 \, \texttt{MeV}$.
We show a vertical line at this time to denote an imagined noise barrier for our chiral computation. 
Fitting the nucleon two-point function with a single 
exponential form at 
$m_\pi = 300 \, \texttt{MeV}$ for 
$6 \leq \tau / a \leq 12$ 
will overestimate the nucleon mass by 
$\sim 3 \%$. 
For the plot at the physical pion mass, 
we assume that the imagined noise barrier remains at 
$\tau / a = 12$. 
Fitting the two-point function on a smaller time interval, 
say $8 \leq \tau / a \leq 12$ results in the same 
quantitative overestimation of the nucleon mass. 
We can see, moreover, 
that improving the signal-to-noise at light pion masses will expose effective masses which continue to fall with time. 
These qualitative features can be deduced directly from the positivity of the spectral representation.

One final observation about the nucleon two-point function is as follows. 
Improving the overlap with the nucleon in two-point functions largely 
removes contamination from the pion-nucleon continuum, 
as evidenced by Figure~\ref{f:Mass}. 
Due to the exponential suppression, 
pion-delta contributions are largely unaffected by the improvement. 
Such contributions are not necessarily suppressed, however, in other nucleon correlation functions.
The axial correlator provides such an example.

\section{Axial Three-Point Function}                       %
\label{s:ThreePoint}                                                 %

We now treat the nucleon axial-current correlation function at finite times using the coordinate-space approach.
Formally there are no spectral representations for three-point functions. 
Consequently contamination from excited states can drive the correlator up or down depending on the underlying dynamics.

\subsection{Chiral Computation}

Nucleon three-point functions are constructed using the LSZ reduction formula to isolate
single-nucleon contributions from the external legs. 
We are interested in a three-point function formed from inserting the axial current
$J^+_{5 \mu} (\bm{y},t)$ 
between two nucleon states. 
Following the typical lattice procedure (for recent lattice calculations of the axial charge, 
see~\cite{Edwards:2005ym,Khan:2006de,Lin:2008uz,Yamazaki:2008py,Alexandrou:2008rp}),
and projecting both source and sink onto vanishing three momentum, 
we form the ratio 
\begin{equation}
\cR_{5 \mu} (\tau, t)
=
\frac{
\int d \bm{x} \int  d \bm{y} \,
\langle 0 | 
N(\bm{x}, \tau ) J^+_{ 5 \mu} (\bm{y}, t) N^\dagger (\bm{0}, 0)
| 0 \rangle 
}
{
\int d \bm{x} \,
\langle 0 | 
N(\bm{x}, \tau )  N^\dagger (\bm{0}, 0)
| 0 \rangle 
}
.\end{equation}
Using the axial current at tree-level in the effective theory, 
namely 
$J^+_{5\mu} 
= 
2 g_A \,  
N^\dagger S_\mu \tau^+ N$, 
we arrive at
\begin{eqnarray}
\cR_{5 \mu} (\tau, t)
=
\theta ( \tau - t ) \theta (t) 
\,
2 g_A  
\, 
u^\dagger S_\mu u
,\end{eqnarray}
where $u$ is a Pauli spinor. 
Working beyond tree level, there are divergences. 
After regularization, 
we renormalize to the physical axial charge,
$G_A$, 
namely
\begin{eqnarray}
\cR_{5 \mu} (\tau, t)
&\overset{\tau \gg t \gg 1}{\longrightarrow}&
2 G_A  \, 
u^\dagger S_\mu u
,\end{eqnarray}
which additionally absorbs the pion-mass dependence of the axial charge into the physical coupling 
$G_A$.

%
\begin{figure}
\epsfig{file=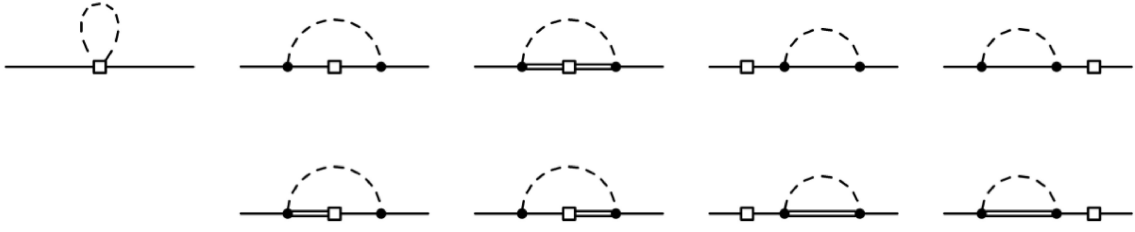,angle=270,width=15cm}
\caption{\label{f:A3pt} 
One-loop diagrams contributing to the axial-vector current matrix element of the nucleon. 
The open squares denote axial-vector current insertions, while the remaining diagram
elements are as in Figure~\ref{f:NM}. 
}%
        \end{figure}
%

The leading one-loop diagrams contributing to the three-point function of the 
axial current are shown in Figure~\ref{f:A3pt}. 
The required terms of the baryon axial current appear as~\cite{Jenkins:1990jv,Jenkins:1991es}
\begin{equation}
J^+_{5\mu} 
= 
2 g_A \,  
N^\dagger S_\mu \tau^+ N
+
g_{\D N}
\left(
T^\dagger_\mu \tau^+ N
+ 
N^\dagger \tau^+ T_\mu
\right)
-
2 g_{\D\D}
\bm{T}^\dagger S_\mu \tau^+ \cdot \bm{T}
,\end{equation}
while the pion-nucleon and pion-nucleon-delta interaction terms 
are contained in the Lagrangian Eq.~\eqref{eq:Lint}. 
Additionally to calculate the ratio 
$\cR_{5\mu}(\tau,t)$, 
we must divide by the two-point function calculated to one-loop order. 
The result of the chiral computation can be cast in the form
\begin{equation}
\cR_{5\mu}(\tau,t)
=
2 u^\dagger S_\mu u \,
\Big[
G_A 
+ 
F_A(\tau,t)
\Big]
\theta ( \tau - t ) \theta (t) 
,\end{equation}
where the function 
$F_A(\tau,t)$ 
encodes the deviation from the limit of infinite time separation. 
The form of this function is
\begin{eqnarray}
F_A(\tau,t) 
&=&
\frac{8 g_A}{9}
\int_{m_\pi}^\infty dE \, \ol \rho_{N \pi}(E)
\left[   e^{ - E ( \tau - t) } + e^{ - E t}  -  e^{ - E \tau} \right]
\notag \\
&&
-
\frac{16 g_{\D N} }{27 \sqrt{2}}
\int_{m_\pi}^\infty dE \, 
\sqrt{ \ol \rho_{N \pi}(E) \ol \rho_{\D \pi}(E+\D)}
\Big[
e^{ - E ( \tau - t)}
\left( 
1 + e^{ - \D ( \tau - t)}
\right)
\notag \\
&& \phantom{spacedout}
+
e^{- E t}
\left(
1 + e^{ - \D t}
\right)
-
e^{ - E \tau}
\left( 
e^{- \D t} + e^{ - \D ( \tau - t)}
\right)
\Big]
\notag \\
&&
+
\left( g_A +  \frac{25 g_{\D\D} }{81} \right)
\int_{m_\pi + \D}^\infty dE \, \ol \rho_{\D \pi} (E)
\left[
e^{ - E ( \tau - t) } + e^{ - E t} 
-
e^{ - E \tau}  
\right] \label{eq:RAxial}
.\end{eqnarray}
The nucleon-pion, $\ol \rho_{N\pi} (E)$, and delta-pion, $\ol \rho_{\D\pi} (E)$,
spectral weights have been given above in Eqs.~\eqref{eq:Npi} and \eqref{eq:Dpi}. 
In the limit 
$\{ \tau, t \} \to \infty$ 
with 
$\tau > t$, 
accordingly we have 
$F_A(\tau, t) \to 0$.

\subsection{Axial Charge Extraction}                           %

\begin{figure}
\epsfig{file=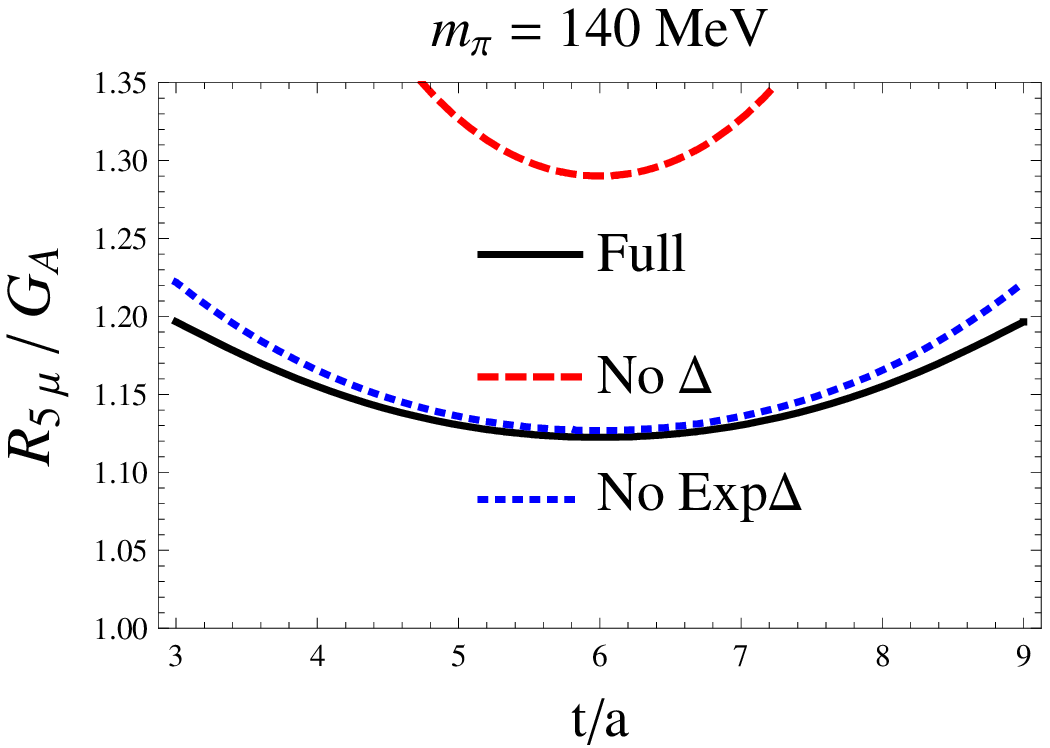,width=7.5cm}
$\quad$
\epsfig{file=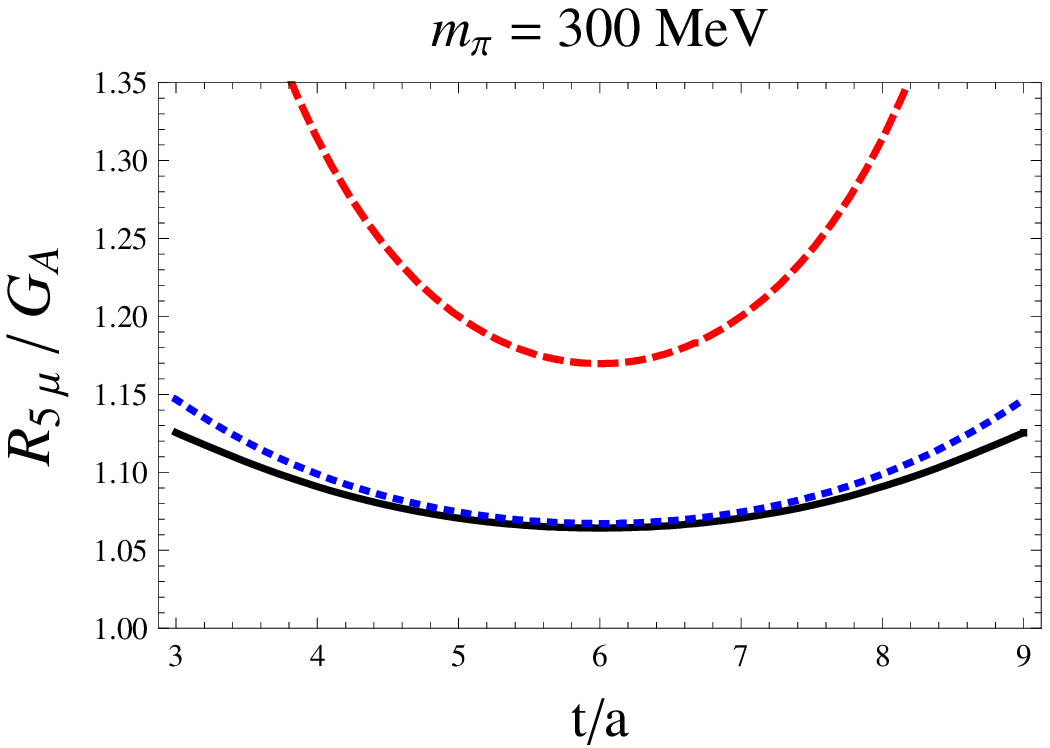,width=7.5cm}
\caption{\label{f:GA} 
Time-dependence of the axial current correlation function.
For sink time $\tau / a = 12$, 
we plot the ratio of the axial three-point to two-point correlators
as a function of the current insertion time $t / a$. 
This ratio we scale by the axial charge $G_A$. 
The curve denoted by \emph{Full} corresponds to the 
axial correlation function determined from Eq.~\eqref{eq:RAxial}, 
while \emph{No $\D$} corresponds to the correlator without
delta-resonance contributions, i.e.~$g_{\D N} = 0$. 
The curve \emph{No Exp$\D$} excludes from the axial 
correlator contributions that are suppressed 
by an exponential factor involving the mass splitting $\D$,
see Eq.~\eqref{eq:NoExpD}. 
}%
        \end{figure}

To investigate the effect of finite times
on the extraction of the nucleon axial charge, we plot
the ratio of axial three-point to two-point functions, 
$R_{5\mu} (\tau,t)$,
given simply by
\begin{equation}
R_{5\mu}(\tau,t) = G_A + F_A(\tau,t)
.\end{equation}
Scaling  $R_{5\mu} (\tau,t)$ by the axial charge, $G_A$, we 
thus expect a plateau at unity for asymptotically
large time separations, $\tau \gg t \gg 1$. 
We fix the sink time at $\tau / a = 12$, 
which is a typical source-sink separation in lattice QCD computations of three-point functions. 
With the sink time $\tau$ fixed, the current insertion time dependence of $R_{5\mu}$ is shown in 
Figure~\ref{f:GA}.
We compare the behavior with and without 
delta resonances. 
Including the resonances requires knowledge of the
delta axial charge, $g_{\D\D}$. 
This low-energy constant is poorly known, see~\cite{Jiang:2008we}, 
and we assume the value $g_{\D \D} = - 2.25$. 
Rather fortuitously the time-dependence of the 
axial correlator is largely insensitive to the value
of the delta axial charge. 
To a good approximation, we can drop all terms in 
Eq.~\eqref{eq:RAxial} that arise from diagrams 
with only intermediate state deltas, that is
\begin{eqnarray}
F_A(\tau,t) 
&\approx&
\frac{8 g_A}{9}
\int_{m_\pi}^\infty dE \, \ol \rho_{N \pi}(E)
\left[   e^{ - E ( \tau - t) } + e^{ - E t}  -  e^{ - E \tau} \right]
\notag \\
&&
-
\frac{16 g_{\D N} }{27 \sqrt{2}}
\int_{m_\pi}^\infty dE \, 
\sqrt{ \ol \rho_{N \pi}(E) \ol \rho_{\D \pi}(E+\D)}
\Big[
e^{ - E ( \tau - t)}
\left( 
1 + e^{ - \D ( \tau - t)}
\right)
\notag \\
&& \phantom{spacedout}
+
e^{- E t}
\left(
1 + e^{ - \D t}
\right)
-
e^{ - E \tau}
\left( 
e^{- \D t} + e^{ - \D ( \tau - t)}
\right)
\Big]
\label{eq:NoExpD}
,\end{eqnarray}
which is independent of $g_{\D\D}$. 
The figure confirms that Eq.~\eqref{eq:NoExpD}
well approximates the time-dependence of the axial correlation 
function. 
Excluding the delta completely (by setting $g_{\D N} = 0$), 
however, 
has a dramatic effect on the time-dependence of the axial correlation function. 
The figure shows that the contributions from pion-nucleon excited states
drive the axial correlator upward. 
Those from pion-delta states are negligible, 
while contributions from the interference of pion-delta and pion-nucleon states
drive the axial correlator downward.

%
%

We see the effect of insufficient time
can lead to spurious plateaus for three-point functions. 
At both pion masses depicted, 
the derived correlation function flattens out, 
but above the asymptotic value of unity. 
Our expectation from chiral perturbation theory
is that the nucleon axial charge would be overestimated, 
with the overestimation worsening as the pion mass is lowered. 
Rather large couplings of the pion to the nucleon, and 
nucleon-delta transition complicate extraction
of quantities from lattice three-point functions. 
As such systematic errors, moreover, depend on the pion mass, 
the resulting chiral behavior of 
$G_A$, 
for example, 
will be specious. 
Lattice QCD calculations often
extract the ratio of the axial-to-vector charges, 
$G_A / G_V$.
The vector charge, 
$G_V$, 
is determined from the three-point function of the vector current which is also susceptible
to an effect from insufficient time. 
The one-loop calculation of the vector three-point function 
in heavy baryon chiral perturbation theory, 
however, 
does not produce any finite time corrections
due to exact cancellation borne in by the vector Ward identity. 
This cancellation is only guaranteed at zero momentum transfer.
Thus away from this limit,  the vector three-point function too 
is subject to finite-time corrections.

As a final comment, 
there has been debate about underestimation of the axial charge in lattice computations~\cite{Jaffe:2001eb,Cohen:2001bg}.
The most recent study of the axial charge
in fully dynamical three-flavor simulations~\cite{Yamazaki:2008py}
finds 
$G_A$ 
falling as the pion mass is decreased.
This behavior was attributed to a finite volume effect,
but not that described by effective field theory~%
\cite{Beane:2004rf,Detmold:2004ap,Smigielski:2007pe}.%
\footnote{
While the volume dependence employed in~\cite{Yamazaki:2008py}
is inspired by the $p$-regime of chiral perturbation theory, 
the size of the corrections is not consistent with expectations~\cite{Beane:2004rf}. 
}
We undertook the present study with the hope that the falling value of 
$G_A$
could be attributed to excited state contamination: 
i.e.~%
although the axial charge is calculated in~\cite{Yamazaki:2008py} on two volumes, 
these volumes have differing temporal extents, 
and hence differing effects from insufficient source-sink separation. 
Given our results, 
there is a plausible way in which such underestimation is possible.
Underestimation of the axial correlation function can arise from 
improvement of the nucleon overlap in two-point functions.
Designing baryon operators with the best ground-state overlap in a two-point function does not necessarily optimally improve three-point functions. 
In a two-point function, 
improvement of the the nucleon operator decreases contamination from the pion-nucleon continuum, 
but not from the the pion-delta continuum,
because
the latter makes only negligible contributions. 
In the axial three-point function, 
however,
the dominate excited state contributions arise from 
both the square of the overlap with pion-nucleon states,
and the interference term between pion-nucleon and pion-delta states. 
Improving the nucleon operator diminishes the former faster than the latter. 
As the latter is negative, 
the axial correlator can thus be underestimated for 
highly optimized nucleon interpolating fields.

\section{Summary}                                                     %
\label{s:summy}                                                          %

In an interacting quantum field theory, 
single particle states emerge only in the limit of long times. 
The existence of particle interactions necessarily 
implies that at shorter times a single particle state can fluctuate into 
multiparticle configurations. 
Such configurations can affect the extraction of hadronic observables.
In particular, 
the existence of light pions in the spectrum of QCD complicates the extraction of hadronic properties as they introduce
a characteristically long time scale,
$\tau \sim 1 / m_\pi$, 
over which pion branch-cuts can be important. 
Using the nucleon in chiral perturbation theory as an example, 
we investigated multiparticle contributions to two- and three-point functions. 
Specifically we found the nucleon mass is subject to overestimation, 
which is directly related to the positivity of the spectral function. 
For the nucleon axial charge, 
$G_A$,
we found that three-point functions calculated away from asymptotic time separations 
are affected by the competition between pion-nucleon and pion-delta contributions. 
Pion-nulceon excited states drive the axial correlation function upward, 
while the interference between pion-delta and pion nucleon excited states 
drives the axial correlation function downward. 
The sign of these corrections depends on the underlying chiral dynamics, 
and cannot be deduced from general principles alone.

The formulae derived in this work apply to the renormalized nucleon operator in chiral perturbation theory. 
Because high-lying excitations are absent, 
this operator behaves like a quark-smeared lattice QCD interpolating field. 
While not quantitative, 
the analytic insight offered by our computation, 
however, 
is useful for comparing qualitatively with lattice QCD computations, 
nearly all of which use optimized baryon operators with smeared quark fields. 
Using a basis of interpolating operators, 
a variational method  
can be employed to isolate ground states from contamination by excited states~\cite{Michael:1985ne,Luscher:1990ck}.
Work along these lines has been pursued,  
for example, 
the baryon spectrum has been investigated in%
~\cite{Burch:2006cc,Basak:2007kj,Lin:2008pr,Bulava:2009jb}. 
Removing pion-nucleon contamination from the nucleon 
two-point function should improve the mass determination, 
although the study in~\cite{Lin:2008pr} seems to overestimate the nucleon mass. 
Despite eliminating a large part of the excited-state contamination, 
the nucleon effective mass continues to drop 
(their Figure~6 has the same qualitative behavior as our Figure~\ref{f:Mass}),
signaling the need for a two-exponential fit. 
Only a single exponential fit, however, was carried out.

Our calculation of the nucleon axial current 
makes plausible that the use of improved nucleon operators can lead to 
an underestimation of the axial charge. 
The extraction of nucleon observables from three-point functions is 
potentially fallible due to the pion continuum, 
which is important owing to the lightness of the pion, 
as well as the comparatively large axial coupling, 
$g_A$. 
The same observation can be made for three-point functions
involving deltas due to the size of the axial transition coupling,
$g_{\D N}$.
As pion masses approach the physical point, 
the effect of insufficient time in lattice QCD calculations of nucleon and delta properties will become pronounced. 
Optimizing the overlap with the ground state in two-point functions
will not necessarily lead to better determination of observables from three-point functions
because of interference terms. 
We hope our work can qualitatively guide the extraction of nucleon and delta observables from lattice QCD.

%

\begin{acknowledgments}
We acknowledge the Institute for Nuclear Theory for their hospitality,
and support from the
U.S.~Department of Energy,
under
Grant No.~DE-FG02-93ER-40762.
\end{acknowledgments}

\appendix

\bibliography{hb}

\end{document}